\documentclass[]{spie}  %>>> use for US letter paper
\usepackage[]{graphicx}
\usepackage{footnote} %for savenotes and minipage
\makesavenoteenv{tabular}
\usepackage{wrapfig} %for wrapfig

\title{Swimming with ShARCS: Comparison of On-sky Sensitivity With Model Predictions for ShaneAO on the Lick Observatory 3-meter Telescope} 

\def\ni{\noindent}

\providecommand{\e}[1]{\ensuremath{\times 10^{#1}}}
\providecommand{\tr}[1]{\textrm{#1}}

\author{
Srikar Srinath\supit{a}, Rosalie McGurk\supit{a}, Constance Rockosi\supit{a}, Renate Kupke\supit{b}, Donald Gavel\supit{b}, Gerald Cabak\supit{b}, David Cowley\supit{b}, Michael Peck\supit{b}, Christopher Ratliff\supit{b}, Elinor Gates\supit{b}, Michael Peck\supit{b}, Daren Dillon\supit{b}, Andrew Norton\supit{b}, Marc Reining\supit{b}
\skiplinehalf
\supit{a}Department of Astronomy \& Astrophysics, University of California, Santa Cruz, CA 95064, USA \\
\supit{b}U.C. Observatories Laboratory for Adaptive Optics, 1156 High St. Santa Cruz, CA 95064
}

\authorinfo{Further author information: (Send correspondence to S.S.)\\
		    S.S.: e-mail: ssrinath@ucsc.edu, telephone: 1 831 459 3068}
\begin{document} 
  \maketitle 

%%%%%%%%%%%%%%%%%%%%%%%%%%%%%%%%%%%%%%%%%%%%%%%%%%%%%%%%%%%%% 
\begin{abstract}
The Lick Observatory's Shane 3-meter telescope has been upgraded with a new infrared instrument (ShARCS - Shane Adaptive optics infraRed Camera and Spectrograph) and dual-deformable mirror adaptive optics (AO) system (ShaneAO). We present first-light measurements of imaging sensitivity in the Ks band. We compare measured results to predicted signal-to-noise ratio and magnitude limits from modeling the emissivity and throughput of ShaneAO and ShARCS. The model was validated by comparing its results to the Keck telescope adaptive optics system model and then by estimating the sky background and limiting magnitudes for IRCAL, the previous infra-red detector on the Shane telescope, and comparing to measured, published results. We predict that the ShaneAO system will measure lower sky backgrounds and achieve 20\% higher throughput across the $JHK$ bands despite having more optical surfaces than the current system. It will enable imaging of fainter objects (by 1-2 magnitudes) and will be faster to reach a fiducial signal-to-noise ratio by a factor of 10-13. We highlight the improvements in performance over the previous AO system and its camera, IRCAL.
\end{abstract}

%>>>> Include a list of keywords after the abstract 


%%%%%%%%%%%%%%%%%%%%%%%%%%%%%%%%%%%%%%%%%%%%%%%%%%%%%%%%%%%%%
\section{INTRODUCTION}
\label{sec:intro}  % \label{} allows reference to this section
The use of laser guide star (LGS) adaptive optics (AO) for astronomical observations was pioneered at Lick Observatory in the mid-90s \cite{maxsci} on the Shane 3-meter telescope. The previous infrared camera, IRCAL, was installed and characterized in 2000 \cite{lloyd}. A major system upgrade, called ShaneAO, has been installed, tested \cite{gavel14} and has seen first light on the Shane telescope \cite{mcgurk}. This upgrade comprises:
\begin{itemize}
\item A new infra-red camera (called ShARCS -- Shane Ao infraRed Camera and Spectrograph) based on the Teledyne HAWAII-2RG detector that has higher resolution, higher quantum efficiency and lower noise than the PICNIC array used in IRCAL \cite{kupke}. The change in detector alone enables longer exposure times before detector limits are reached (where night sky lines are not dominant).

\item A dual-deformable mirror AO pipeline with a 52-actuator, high-stroke, low-frequency (spatial and temporal) ALPAO DM-52 mirror (the ``woofer'') and a 1024-actuator, low-stroke, high-frequency MEMS Boston Micromachines Kilo-DM mirror (the ``tweeter'') \cite{norton}. The Shack-Hartmann wavefront sensor camera has been upgraded to allow use of a 30$\times$30 lenslet array.

\item A more efficient solid-state laser \cite{dawson} (replacing the current dye laser) with projected improvements to the launch system geared towards increasing return flux so that the AO system can use more subapertures (not yet installed). 

\item A new support structure for the AO system and camera designed to reduce flexure and improve long-exposure stability \cite{ratliff}. The assembly is also designed to be rotated with precision to allow for long-exposure, long-slit spectroscopy.
\end{itemize}

The cumulative effect of the upgrades is diffraction-limited imaging and Nyquist-sampled point-spread functions in the $J$, $H$ and $K$ bands \cite{kupke} at nearly double the Strehl ratio of the current system \cite{gavel11}. Nyquist-sampling in all bands is particularly desirable to measure and reconstruct the PSF reliably. The PSF varies because of atmospheric turbulence and the changing gravity vector as the telescope moves -- the AO bench is mounted at the Cassegrain focus. 

The new laser with a pulse format designed to couple better to the Sodium atoms in the upper atmosphere is expected to allow routine use of more subapertures, making better use of the tweeter's spatial frequency sampling ability. Hence, any turbulence will be better measured and corrected because subaperture spacing will be 10 cm for a 30$\times$30 lenslet array, which is a much better match to measured conditions at Lick Observatory \cite{gavel03}. In addition, the telescope will be usable over a greater range of the nightly and seasonal variations in mesospheric Sodium levels. 

Spectroscopy will also be much improved because of the aforementioned improvements in stability. The ability to reliably rotate the instrument structure and the slit will allow stable object tracking over a longer course of time. Less object wander enables reliable co-adding to raise signal-to-noise ratio (SNR), better tracking of variability in night sky lines and minimizes the impact of hot pixels. The intent of these upgrades was to investigate fainter objects at the diffraction limit of the telescope in all bands and achieve target SNRs in less time. We present the result of modeling the emissivity and throughput of ShaneAO and conservatively predict that the new system will achieve its targets based on first-light measurements. We start by defining and justifying our choice of sky and telescope model in Section \ref{sec:sky}. The model is validated by reproducing predicted sensitivity for the NIRC2 instrument on the Keck telescope and then by checking whether it can account for observed sky backgrounds and throughput for the earlier IRCAL system on the Shane telescope in Section \ref{sec:validation}. We present a comparison of expected photons received as predicted by the model and as measured by observing a spectrophotometric standard in Section \ref{sec:first light}.

%%%%%%%%%%%%%%%%%%%%%%%%%%%%%%%%%%%%%%%%%%%%%%%%%%%%%%%%%%%%%
\section{SKY MODEL}
\label{sec:sky}  % \label{} allows reference to this section

%%-------------
   \begin{figure}[t]
   \begin{center}
   \begin{tabular}{c}
   \includegraphics[scale=0.65]{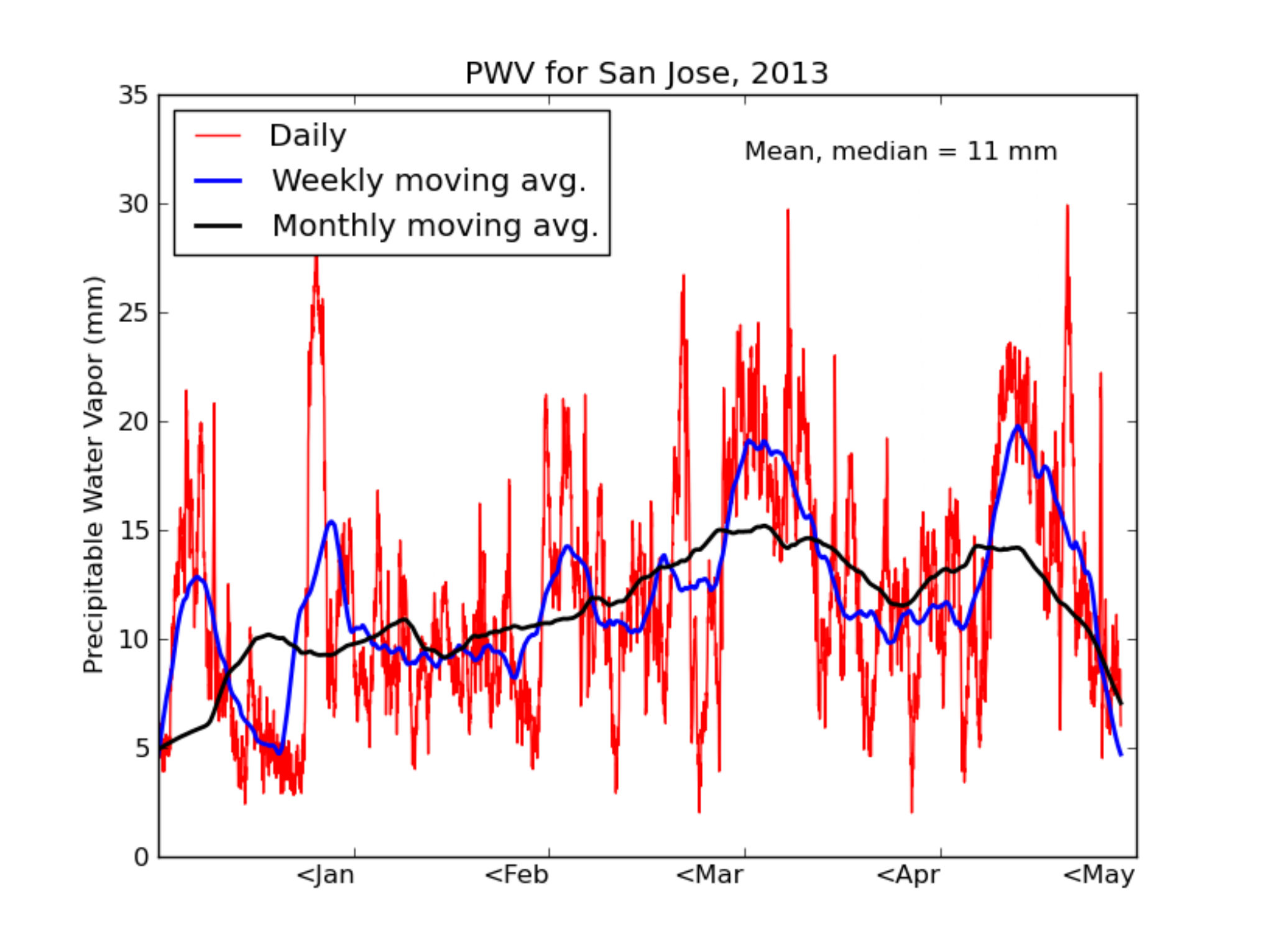}
   \end{tabular}
   \end{center}
   \caption
   { \label{fig:pwv} 
	Precipitable water vapor (PWV) above San Jose airport (altitude: sea-level) for the first half of 2013. Mt. Hamilton is higher (1284 m) and drier. Based on the median PWV of 11 mm for San Jose, the choice of 10 mm as a representative PWV for our sky model above Mt. Hamilton is justifiable. The time period displayed spans Winter, Spring and Summer. Fall conditions are very similar to Spring.}
   \end{figure} 
%%-------------

In the near infrared (NIR) band (1-2.3 $\mu$m), the main source of emission in the night sky is from $OH^-$ lines in the mesosphere \cite{meinel}. Beyond 2.3 $\mu$m, $O_2$ and $H_2O$ lines situated at lower altitudes dominate. However, at these wavelengths thermal emission from the telescope and optics overwhelms any atmospheric emission \cite{ramsay}. Non-thermal emission from the NIR sky varies on timescales of 10-15 minutes by 10\% or more making it difficult to estimate a nominal value for the background. 

The water column above a site is another variable in the sky spectrum. Weather data from the San Jose, CA, airport weather station for the first half of 2013\footnote{available from http://www.suominet.ucar.edu/index.html} is plotted in Figure \ref{fig:pwv}. The mean and median precipitable water vapor for that site is $\sim$11 mm. Lick Observatory is situated on Mt. Hamilton at an altitude of 1284 m. This is a higher and drier site than the airport so assuming a water column value of 10 mm is a reasonable estimate.

\begin{figure}[h]
   \begin{center}
   \begin{tabular}{c}
   \includegraphics[scale=0.65]{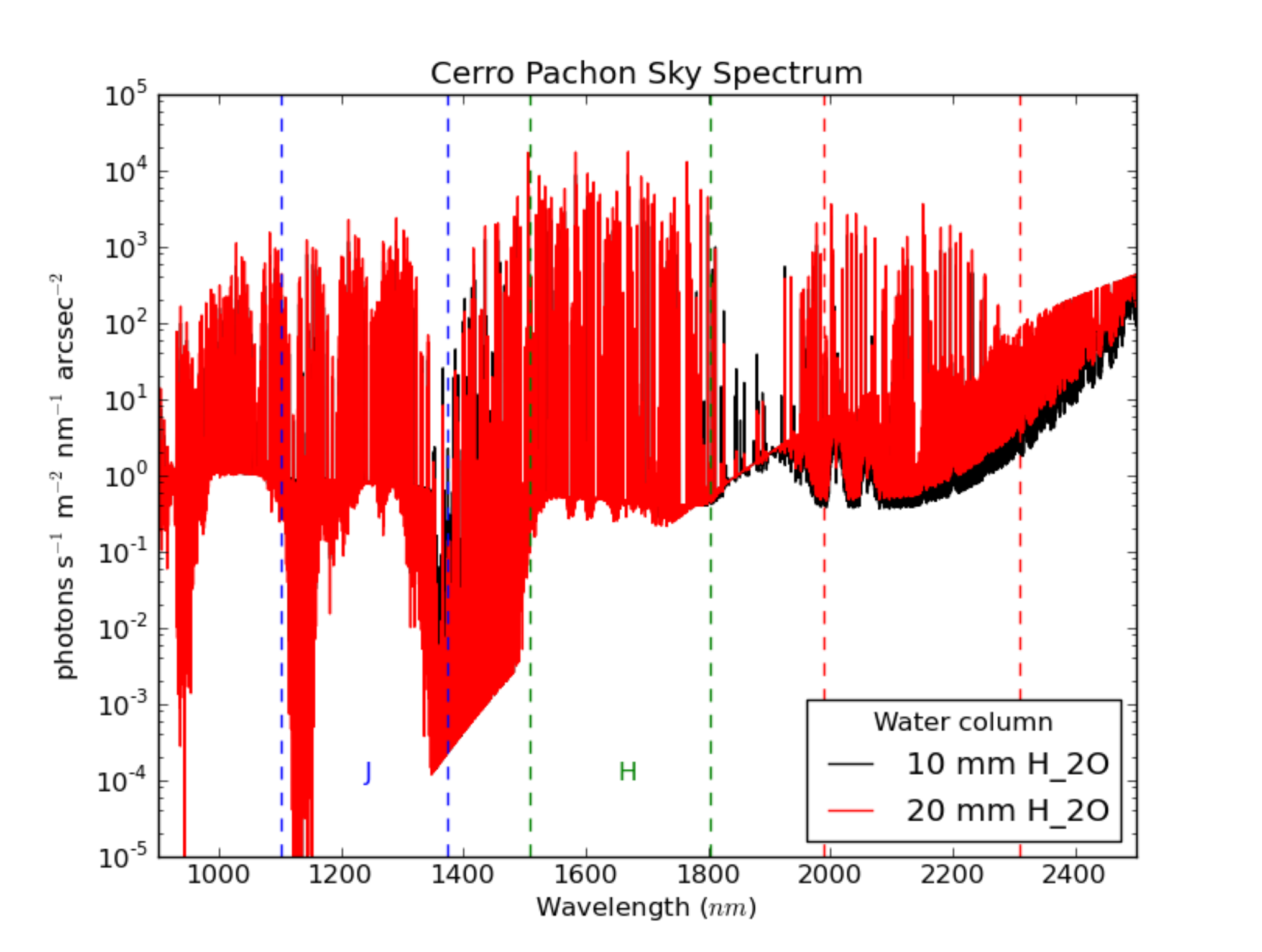}
   \end{tabular}
   \end{center}
   \caption
   { \label{fig:h2o} 
	Sky background in the NIR for 10 mm (black) and 20 mm (red) water columns. The 20 mm curve is equivalent to looking through an airmass of 2.0 for a 10 mm water column. Significant differences in the two curves are visible only in the longest wavelength end (K-band) where telescope emissivity is dominant.}
   \end{figure}

For our purposes we chose sky emission and transmission models available on the Gemini Observatory website\footnote{at http://www.gemini.edu/?q=node/10787} for Cerro Pachon (altitude 2700 m) with an airmass of 1.0, water column of 10 mm and with a spectral resolution of $R\sim 2000$. The spectra are constructed with high-resolution sky transmission data generated by ATRAN \cite{lord} to which a 280 K blackbody spectrum for sky emission, OH and O$_2$ lines, and zodiacal light have been added. The effects of moonlight (which would be strongest in the $J$ band) have not been accounted for. Water column and airmass make a difference in the $K$ band as seen is Fig \ref{fig:h2o}, where telescope emission dominates so the predicted magnitude limits for ShaneAO should hold for water columns up to 20 mm or an airmass of 2.0 for a 10 mm water column.

%%-------------
%%-------------

The telescope mirrors, AO and instrument optics are modeled as a series of gray bodies, each with emissivity $\epsilon = 1 - \tau$, where $\tau$ is the transmissivity/reflectivity of a surface. Transmission and reflection curves were either measured or supplied by manufacturers (such as ALPAO for the woofer). For surfaces without data, a flat curve was assumed across the wavelength band based on reasonable assumptions. 

The effects of dust or degraded coatings, both a real problem at lower altitude sites like Lick Observatory, were modeled by modifying the emissivity or transmissivity of a surface as follows:

\begin{itemize}
\item Degraded coatings (like Aluminum on the primary and secondary mirrors) had their reflectivity or transmissivity uniformly cut by 2.5\%
\item A variable dust fraction, $f_D$, with emissivity $\epsilon_D=1$ (i.e. dust was modeled as a perfect blackbody) was used to change emissivity:
	\begin{eqnarray}
		\epsilon_{eff} &=& (1 - f_D)\epsilon + f_D\epsilon_D \\
		\tau_{eff} &=& 1 - \epsilon_{eff}
	\end{eqnarray}
\end{itemize}

The telescope and AO system were set at a worst case ambient temperature of 18$^\circ$ C and the change in emissivity due to dust is applied to these surfaces. Instrument optics, filter wheels and the IR detector reside in a dewar cooled to -196$^\circ$ C and are considered dust-free. 

The sky spectrum described above is propagated through each surface in the system. At any given surface, the upstream flux is reduced by the surface's reflectivity or transmissivity and the dust-modified emissivity of the surface is added. All pupil stops are assumed to behave perfectly for the ShaneAO system, so off-axis contributions to emissivity, from support structures for example, are ignored. 

Hence, at each surface $i$, the flux in photons s$^{-1}$ m$^{-2}$ arcsec$^{-2}$ nm$^{-1}$ at a particular wavelength, $\lambda$, is given by:

\begin{equation}
O_i(\lambda) = O_{i-1}(\lambda)\cdot \tau_{eff_{i}}(\lambda) + \frac{B_\lambda (T_i)}{hc/\lambda} \epsilon_{eff_{i}}(\lambda) \alpha
\end{equation}

where $B_\lambda (T)$ is Planck's blackbody function; $h$, $c$ and $k$ are the usual physical constants; $\alpha$ is a unit conversion factor (it includes conversion from steradians to arcsec$^2$ among others). For a two-temperature system with $N$ surfaces, the flux reaching the detector is given by: 

\begin{equation}
O_N(\lambda) = O_0(\lambda)\cdot \tau_{eff_{1}}(\lambda) \tau_{eff_{2}}(\lambda) + \frac{B_\lambda (T_1)}{hc/\lambda} \epsilon_{eff_{1}}(\lambda)\alpha + \frac{B_\lambda (T_2)}{hc/\lambda} \epsilon_{eff_{2}}(\lambda)\alpha 
\end{equation}

where $\tau_{eff_{i}}$, $\epsilon_{eff_{i}}$ are the effective throughput and emissivity of the parts of the system at temperature $T_i$ (in this model 18$^\circ$ C for the telescope and AO system, -196$^\circ$ C for components in the dewar).

At the detector, we assume diffraction limited imaging in each band and a flat-topped PSF scaled by the Strehl ratio. The collecting area of the telescope is that of the primary mirror with the area obscured by the secondary subtracted. Assuming the gain is set such that the ratio of incident photons to electrons in a well is 1:1, The noise due to the background is then the number of electrons in the Airy core generated by the sky and telescope emissivity in the time spanned by a Fowler-16 (IRCAL) or -32 (ShARCS) read.

%%%%%%%%%%%%%%%%%%%%%%%%%%%%%%%%%%%%%%%%%%%%%%%%%%%%%%%%%%%%%
\section{VALIDATION}
\label{sec:validation}  % \label{} allows reference to this section

\subsection{Comparison with the Keck model}
Results from our model were compared to those for the Keck AO system \cite{bouchez}. The surface list for the telescope and AO system listed in Section \ref{ssec-surfs} and the same sky spectrum, which is a modified version of the Gemini spectrum (airmass 1.5, 1.6 mm H$_2$O), were used. 

\begin{table}[h]
	\centering
		\begin{tabular}{|c|c|c|c|c|}
		\hline
		 Filter & Sky brightness    & dust fraction=0.0 & dust = 0.008    & Avg. Trans\\
		        & mag arcsec$^{-2}$ & mag arcsec$^{-2}$ & mag arcsec$^{-2}$ & dust = 0.008\\
		\hline
		\multicolumn{5}{|l|}{Keck model results} \\
		\hline
		    J   & 16.01 & 15.90 & 15.89 & \\
		\hline
		    H   & 13.78 & 13.72 & 13.71 & \\
		\hline
		    K   & 14.91 & 12.91 & 12.63 & 0.606 \\
		\hline
		\multicolumn{5}{|l|}{ShaneAO model results} \\
		\hline
		    J   & 16.00 & 16.00 & 16.00 & 0.507\\
		\hline
		    H   & 13.80 & 13.80 & 13.80 & 0.544\\
		\hline
		    K   & 14.86 & 13.01 & 12.71 & 0.601\\
		\hline
	\end{tabular}
	\caption{Comparison of Keck AO model results and the ShaneAO model. The first column displays sky brightness as measured at the entrance pupil. The second and third columns show sky brightness as measured at the detector for a dust fraction (percentage area covered by dust) of 0.0 (\textit{i.e.} no dust) and 0.008 or 0.8\% dust coverage. Our model predicts nearly the same sky brightness and is no more than 10\% different for all three bands (J, H and K) considered. Throughput values for the K-band are also very close. Surface list and reflection/transmission curves named ``K2AO'' (Section 5.1\cite{bouchez}). Average transmission in filter is for dust fraction=0.008.}
	\label{tab-keck}
\end{table}

The comparison between models is presented in Table \ref{tab-keck} for dust fractions of 0.0 and 0.8\%. The average transmission in the filter is for a dust fraction of 0.8\%. Sky background is as measured at the entrance pupil of the telescope and at the detector. Our results are no more than 10\% different in flux in the J-band.

\subsection{Comparison to measured IRCAL results}

The background and limiting magnitudes for the preceding camera and AO system were measured and reported \cite{lloyd} and are reproduced in Table \ref{tab-ircal}. The model had to be adjusted to include off-axis emission because the IRCAL cold stop was 18\% oversized. The surface list used for this exercise is in Section \ref{ssec-ircal-surfs}. Strehl ratio for the $K$ band was derived from published measurements\cite{olivier} and scaled for other bands using the relation:

\begin{equation}
S = \tr{exp}\left[-\left(\frac{\lambda_0}{\lambda}\right)^2 \log\left(\frac{1}{S_0}\right)\right]
\end{equation}

where $S_0$ is the reference Strehl (0.42 in LGS mode and 0.65 in NGS mode in the $K$-band) and $\lambda_0$ the central wavelength of the reference filter. Filter curves were derived from those displayed in the IRCAL manual on the Lick Observatory website.\footnote{at http://mthamilton.ucolick.org/techdocs/instruments/ircal/ircal\_filters.html}

\begin{table}[h]
	\centering
		\begin{tabular}{|c|c|c|c|}
		\hline
		 Filter & Gemini sky b/g    & IRCAL sky b/g     & Total\\
		        & mag arcsec$^{-2}$ & mag arcsec$^{-2}$ & Throughput \\
		\hline
		    J   & 16.1 & 16.0 & 0.08\\
		\hline
			H   & 14.3 & 14.4 & 0.16\\
		\hline
			K   & 13.0 &  9.3 & 0.13\\
		\hline
			Ks	& 	   & 10.3 & 0.14\\
		\hline
		\end{tabular}
	\caption{Mt. Hamilton measured sky background by the Gemini instrument
 and IRCAL. IRCAL J-band sky background was not reported initially but estimated later\cite{lloyd}. Total throughput values are for the IRCAL instrument. The values above are from the IRCAL manual on the Lick Observatory website.}
	\label{tab-ircal}
\end{table}

Gemini in Table \ref{tab-ircal} is an older infrared camera which made sky background measurements as well\cite{mclean}. The sky backgrounds predicted by the ShaneAO model with no dust coating the optics and a 3\% dust coating for the existing system (IRCAL) are in Table \ref{tab-ircres}. The predicted values are in good agreement with the measured values in Table \ref{tab-ircal}.

\begin{table}[h]
	\centering
		\begin{tabular}{|c|c|c|c|}
		\hline
		 Filter & dust=0.0        & dust=0.03      & Avg trans\\
		        & mag arcsecs$^{-2}$ & mag arcsecs$^{-2}$ & dust=0.03 \\
		\hline
%		\multicolumn{4}{|l|}{Model results}\\
%		\hline
		    J  & 15.80 & 15.80 & 0.112 \\
		\hline
		    H  & 13.80 & 13.77 & 0.131 \\
		\hline
		    K  &  9.75 &  9.31 & 0.136 \\
		\hline
			Ks & 10.53 & 10.10 & 0.139 \\
		\hline
		\end{tabular}
%	\end{adjustwidth}
	\caption{Mt. Hamilton predicted sky background and throughput numbers by the model of the current Shane AO system with IRCAL. Predictions for the K and Ks bands agree very well with measurements in Table \ref{tab-ircal}. The unusually low measured throughput in the J-band has not been diagnosed.}
	\label{tab-ircres}
\end{table}

Compared to Keck, the dust fraction had to be increased to 3\% to better match the IRCAL system's magnitudes and throughput. Since Mauna Kea (the site for the Keck telescopes) is at $\sim 3.5$ times the altitude of Mt. Hamilton, increasing the dust fraction for IRCAL by a similar factor seems reasonable. Coating degradation (2.5\% for Aluminum and 1\% for other surfaces) is the same for both comparisons. 

Given the variation in NIR sky emission, the numbers in Tables \ref{tab-ircal} and \ref{tab-ircres} are in very good agreement. Further validation is provided by comparing limiting magnitudes for point sources using a natural guide star (NGS) as measured and reported\cite{lloyd} and available online\footnote{from http://astro.berkeley.edu/~jrg/ircal/spie/ircal.html} versus those predicted by our model in Table \ref{tab-irclim}. The unusually low measured throughput in the $J$ band is not reproduced by the model but the decrease in throughput between $J$ and $H$ bands is. 

\begin{table}[h]
	\centering
		\begin{tabular}{|c|c|c|}
		\hline
		 Filter & IRCAL    & IRCAL model    \\
		        & measured & predicted \\
		\hline
		    J   & 21.8 & 21.46 \\
		\hline
		    H   & 20.5 & 20.45 \\
		\hline
		    K   & 17.8 & 18.15 \\
		\hline
			Ks	& 18.3 & 18.34 \\
		\hline
		\end{tabular}
	\caption{Point source limiting magnitudes (Vega system): IRCAL measured limiting point source magnitudes compared with values predicted by our model with IRCAL system surfaces (300s Fowler sample - 16 reads, SNR=5, Natural Guide Star (NGS) mode)}
	\label{tab-irclim}
\end{table}

%%%%%%%%%%%%%%%%%%%%%%%%%%%%%%%%%%%%%%%%%%%%%%%%%%%%%%%%%%%%%

\section{MODEL PREDICTIONS}
\label{sec:modelpredict}  % \label{} allows reference to this section

\subsection{Imaging}
With the new system, we retain the IRCAL filter curves, dust fraction of 3\% and coating degradation described above. The list of surfaces used is in Section \ref{ssec-shaneao-surfs}. Strehl ratios are derived from estimates\cite{gavel11}. The predicted sky background and throughput for ShaneAO are given in Table \ref{tab-NSAOres}. 

\begin{table}[h]
	\centering
		\begin{tabular}{|c|c|c|c|c|}
		\hline
		 Filter & Average sky  & Predicted measured    & Predicted measured & Throughput\\
		 	    & brightness   & brightness (dust=0.0) & brightness (dust=0.03) & (dust=0.03)\\
		        & mag arcsec$^{-2}$  & mag arcsec$^{-2}$ & mag arcsec$^{-2}$ & \\
		\hline
		    J   & 15.84 & 15.80 & 15.80 & 0.136 \\
		\hline
		    H   & 13.85 & 13.84 & 13.82 & 0.175 \\
		\hline
		    K   & 13.87 & 11.84 & 10.40 & 0.165 \\
		\hline
			Ks	& 14.60 & 12.68 & 11.21 & 0.171 \\
		\hline
		\end{tabular}
%	\end{adjustwidth}
	\caption{Sky background and throughput predictions for new ShaneAO system. The first column shows sky brightness at the telescope's entrance pupil. The second and third columns show sky brightness as measured at the detector. Throughput is for a dust fraction of 3\%.}
	\label{tab-NSAOres}
\end{table}

\begin{figure}[h]
   \begin{center}
   \begin{tabular}{c}
   \includegraphics[scale=0.65]{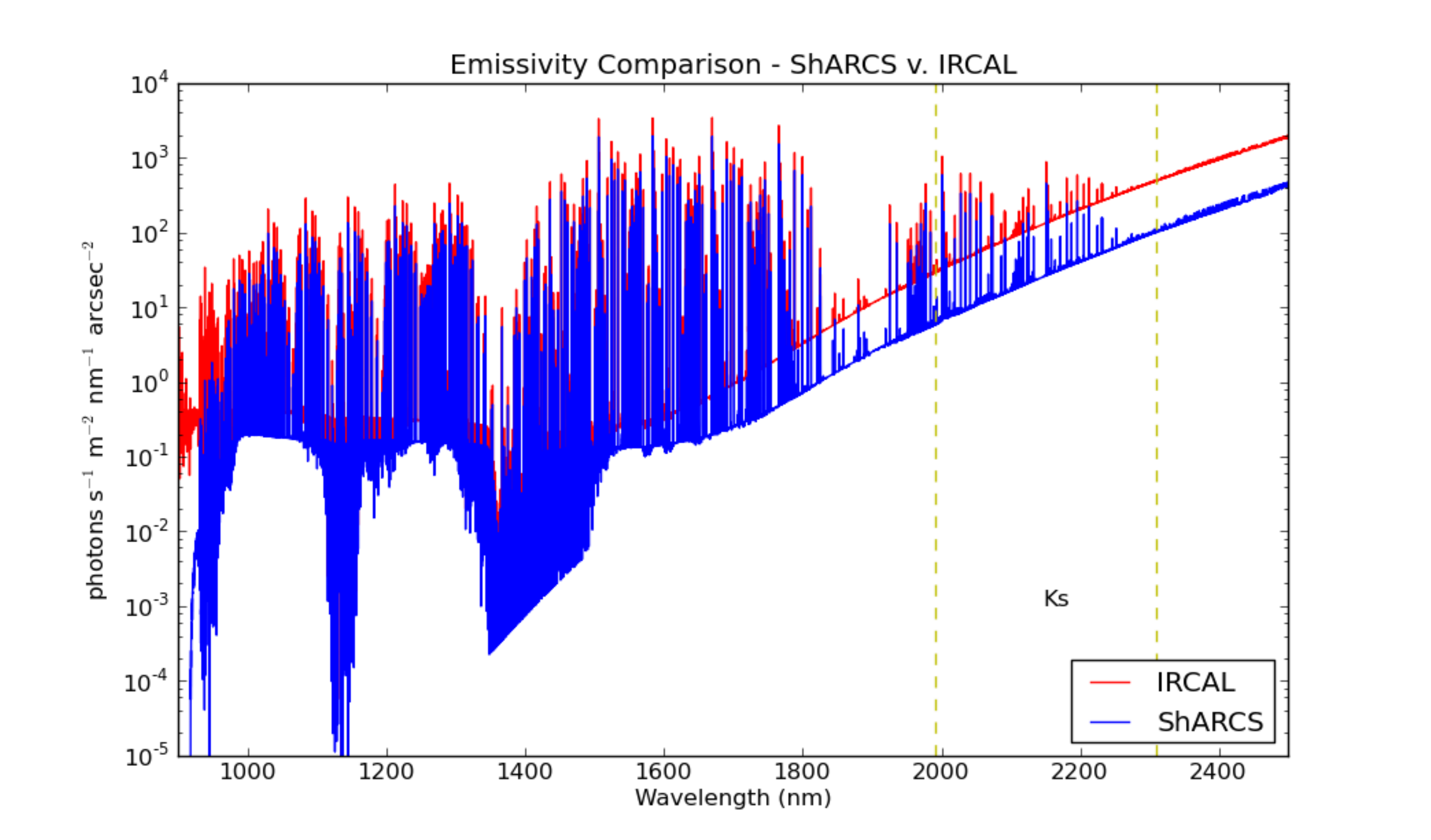}
   \end{tabular}
   \end{center}
   \caption
   { \label{fig-bg} 
	Total background and emissivity comparison as determined by our model between the existing system (IRCAL) and the new system (ShARCS). The new system's background is $10\times$ lower than the older system's, largely because of a properly-sized cold stop eliminating stray, off-axis radiation.}
   \end{figure}

Comparing Tables \ref{tab-NSAOres} and \ref{tab-ircres}, the model predicts that the background in the $K$ band will be a magnitude fainter and throughput will increase by 20-30\% across all bands. The reduced background for ShaneAO is illustrated by Figure \ref{fig-bg}. 

%%-------------
%%-------------

For limiting magnitudes, the comparison between older (IRCAL) and new systems is one of between their usual modes of operation:
\begin{description}
\item[IRCAL:] LGS mode with 8$\times$8 subapertures and 300 second Fowler-16 exposures.
\item[ShARCS+old laser:] LGS mode with 8$\times$8 subapertures (\textbf{ShARCS 8} in the figures and tables below) and 300 second Fowler-16 exposures -- this is the current \textit{modus operandi} until the new laser is operational.
\item[Full ShaneAO:] LGS mode with 16$\times$16 subapertures (\textbf{ShARCS 16} hereafter) and 300 second Fowler-32 exposures (not yet in operation).
\end{description}

\subsection{Point Source Limiting Magnitudes}
Limiting magnitudes for a 5$\sigma$ detection (i.e. SNR=5) of point sources for the older and new systems is shown in Table \ref{tab-lm-lostr}. ShaneAO should be able to image point sources 1.3-2 magnitudes fainter than the existing system can. 

\begin{table}[h]
	\centering
		\begin{tabular}{|c|c|c|c|}
		\hline
		 Filter & IRCAL     & ShARCS 8  & ShARCS 16  \\
		        & predicted & predicted & predicted     \\
		\hline
		K strehl& 0.42 (LGS) & 0.6 (LGS) & 0.8 (LGS) \\
		\hline
		    J   & 19.97 & 21.30 & 22.28 \\
		\hline
		    H   & 19.62 & 20.48 & 21.03 \\
		\hline
		    K   & 17.67 & 18.71 & 19.02 \\
		\hline
			Ks	& 17.84 & 18.92 & 19.24 \\
		\hline
		\end{tabular}
	\caption{Point source limiting magnitudes (Vega system) in LGS mode: IRCAL predicted (300s Fowler sample - 16 reads) and ShARCS (8 or 16 subapertures) predicted (300s Fowler sample - 32 reads). Higher predicted magnitude numbers for ShARCS means greater sensitivity. The values in column 2 for IRCAL differ from those in Table \ref{tab-irclim} because these are limiting magnitudes for LGS mode versus NGS mode in the earlier table.}
	\label{tab-lm-lostr}
\end{table}

\begin{figure}[h]
   \begin{center}
   \begin{tabular}{c}
   \includegraphics[scale=0.65]{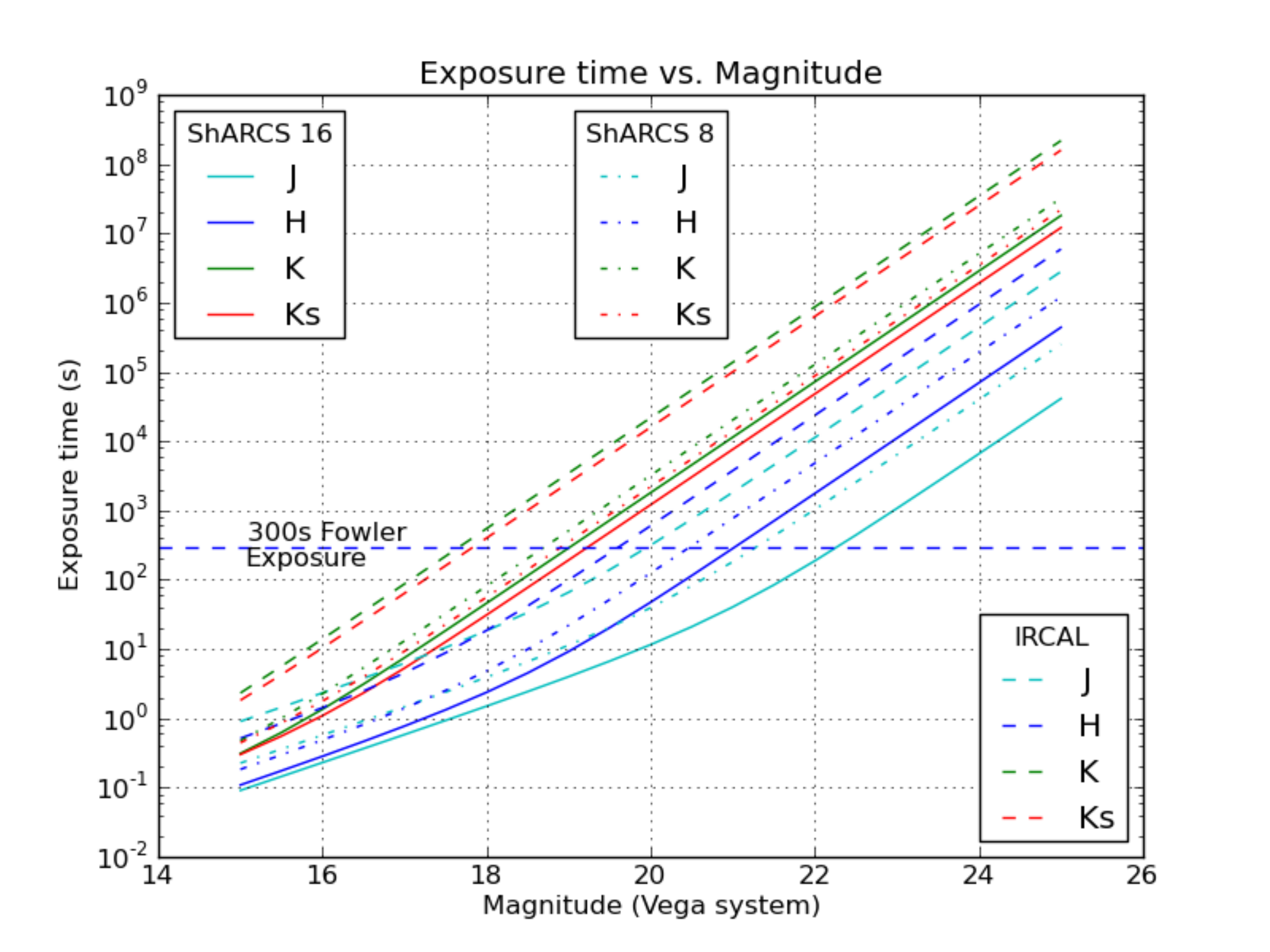}
   \end{tabular}
   \end{center}
   \caption
   { \label{fig-sharcexpt-lostr} 
	Point sources - Exposure time required to reach SNR=5 vs. magnitude (Vega system) in different filters for IRCAL, ShARCS 8 (ShARCS + old laser) and full ShaneAO (ShARCS 16). ShARCS is faster by 10 (ShARCS8) - 13$\times$ (SHARCS16) to reach a fiducial SNR of 5$\sigma$.}
   \end{figure}

\begin{figure}[h]
   \begin{center}
   \begin{tabular}{c}
   \includegraphics[scale=0.65]{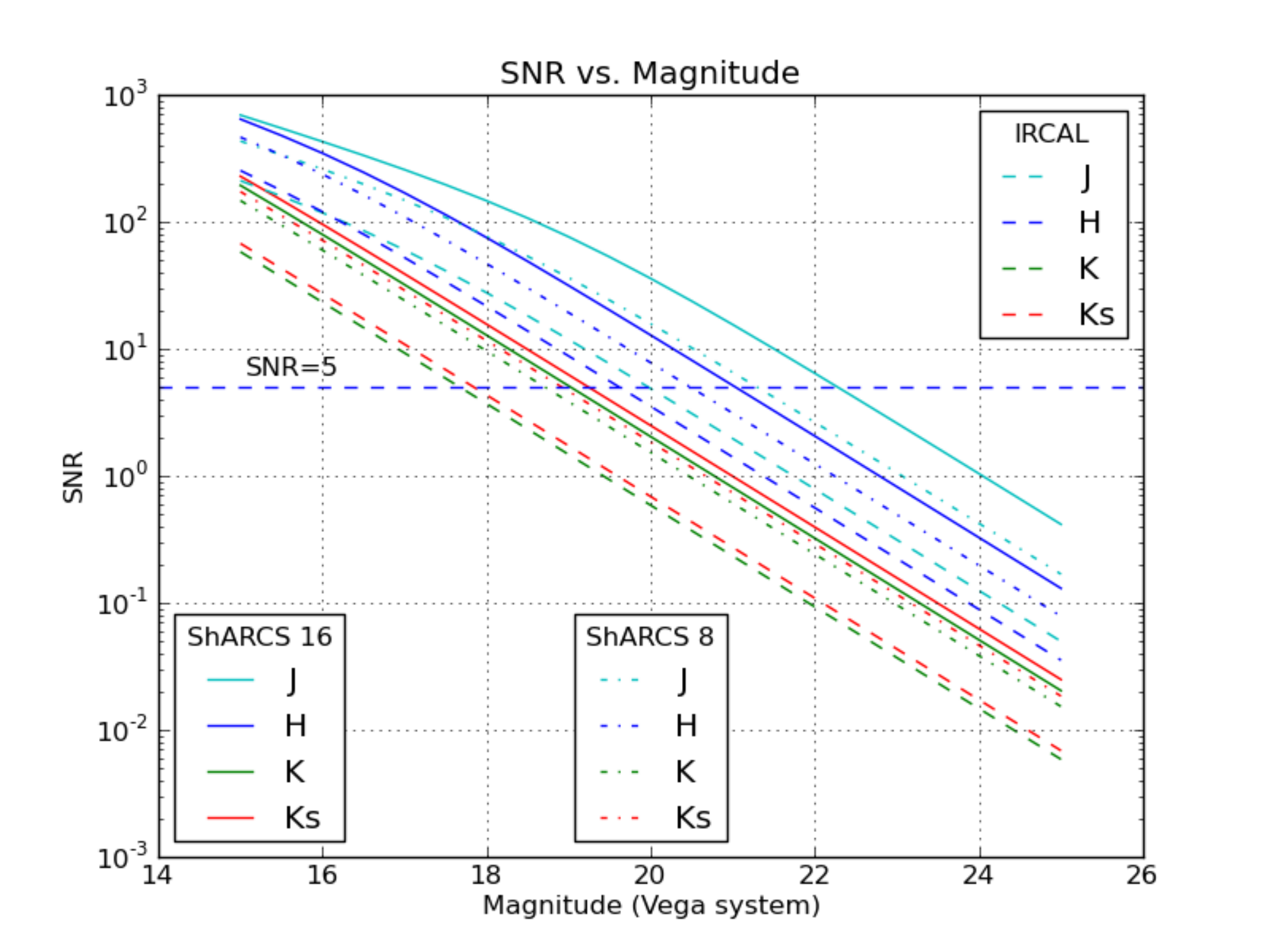}
   \end{tabular}
   \end{center}
   \caption
   { \label{fig-sharcsnr-lostr} 
	Point sources - SNR vs. magnitude for one 300s fowler exposure in different filters for IRCAL and ShARCS (8 or 16 subapertures). ShARCS reaches a fiducial SNR(=5) for much fainter objects than IRCAL did.}
   \end{figure}

Graphically, the results for point sources for older and new systems are displayed in Figures \ref{fig-sharcexpt-lostr} and \ref{fig-sharcsnr-lostr}. Figure \ref{fig-sharcexpt-lostr} illustrates the prediction that ShaneAO will be faster by a factor of 10-13 in reaching a fiducial SNR. Figure \ref{fig-sharcsnr-lostr} is a representation of Table \ref{tab-lm-lostr}. The curves change slope at the transition between sky background noise dominance (fainter sources) and source photon Poisson noise dominance (bright sources). 

%%-------------
%%-------------

%%-------------
%%-------------

%%%%%%%%%%%%%%%%%%%%%%%%%%%%%%%%%%%%%%%%%%%%%%%%%%%%%%%%%%%%%

\section{FIRST LIGHT}
\label{sec:first light}  % \label{} allows reference to this section

A spectrophotometric standard star, BD+26 2606, chosen from the Hubble Space Telescope's (HST) calibration database system\footnote{http://www.stsci.edu/hst/observatory/crds/calspec.html} was imaged in the Ks band and reduced in one of the earliest commissioning runs for ShaneAO/ShARCS. The star's published spectrum (as measured by HST at the top of the atmosphere) was propagated through the ShaneAO sky and system model and the expected number of photons at the detector was compared with measurements. 

Ambient conditions such as a system temperature of 10$^\circ$ C and a Shack-Hartmann subaperture configuration of $16\times 16$ were fed into the model, whereas indeterminate parameters such as coating degradation were held at nominal model values described earlier.  System gain has been measured to be 2.25 electrons per Analog/Digital Unit (ADU) \cite{mcgurk} and we assume an incident photon-electron ratio of 1:1 in the detector. Model-reported photon numbers were converted to ADU for comparison with images. Strehl ratio and dust fraction were maintained as free parameters. The sky photon count was used to constrain dust fraction (since sky counts in the Airy core are independent of Strehl ratio).

A series of 6-second exposures of the star taken on one night was corrected by subtracting dark current from, flat-fielding and cosmic ray-cleaning the images and the number of counts in the diffraction-limited core was logged. No sky subtraction was performed because we want to measure background levels. Background was estimated from an annulus around the point source where the measured counts approached median pixel values for the exposure as a whole. Images were all diffraction-limited in the Ks-band. Sky counts matched best for a dust fraction of 2\% (which is less than assumed for our predictions and arguably justified for a newly-installed instrument). For a Strehl ratio of 0.35 in the model (we are in the process of characterizing a Strehl measuring tool for images, but seeing was not optimal on the night the images were taken and a Strehl ratio of 0.35 is a reasonable estimate), measured and predicted source counts are shown in Table \ref{tab-bd26}.

\begin{table}[h]
	\centering
		\begin{tabular}{|c|c|c|c|}
		\hline
		 Sky counts          & Mean sky counts & Source counts         & Mean source counts \\
		 predicted (df=0.02)   & measured         & predicted (Strehl=0.35) & measured     \\
		\hline
		 7322.6 &  7299.4 & 675075.2 & 665557.7 \\
		\hline
		\end{tabular}
	\caption{Model predictions versus measurements for a spectrophotometric standard star BD+26 2606 propagated through the model and imaged in 6-second expsoures in the Ks-band. All numbers are in units of ADU. First column shows sky counts predicted by the model in the Airy core for a dust fraction of 2\%. The second column shows sky counts in the Airy core based on background levels measured in an annulus around the star image. Using sky values to constrain dust fraction, column 3 shows predicted number of source counts in the Airy core for a Strehl ratio of 0.35 in the Ks-band. Column 4 shows mean measured ADU in the Airy core of a series of 6-second exposures of BD26+2606 in the Ks-band.}
	\label{tab-bd26}
\end{table}

Further work is required to constrain model parameters better and feed it into an exposure time calculator for the ShaneAO system. This work will also be extended to constant surface brightness patches and spectroscopy. Model predictions exist for both types of objects and data is being collected to check those predictions as well.

%%%%%%%%%%%%%%%%%%%%%%%%%%%%%%%%%%%%%%%%%%%%%%%%%%%%%%%%%%%%%
\section{CONCLUSION and FUTURE WORK}
\label{sec:future}  % \label{} allows reference to this section
We have modeled the emissivity and throughput of the new AO system and IR camera installed on the Shane 3-meter telescope at Lick Observatory. The model's predictions fare well when compared with those of the Keck AO system model and the measured background and magnitude limits of the previous AO system and IRCAL detector on the Shane telescope. We conservatively predict that the new system will be able to study fainter objects and reach target SNRs 10-13 times faster than the current system. Analysis of first light images of a spectrophotometric standard star from the new detector show that we are receiving the expected number of photons predicted by the model. Future work will validate model predictions for constant surface brightness patches and for spectroscopy. The model's predictions are predicated on worst-case summertime temperatures, dusty optics (3\% dust fraction) and degraded coatings. On a night with median temperature and seeing, if the optics are kept clean and the primary mirror is re-coated (late 2014), we expect the telescope to perform better than predicted. In addition, we have not yet factored in the longer exposures that should be possible with the stability and tracking improvements afforded by the new instrument assembly. We are confident that the ShaneAO system in conjunction with ShARCS will enable the 3-meter telescope to continue to operate at the frontier of Astronomy as it has done for more than 50 years.

%%-----------------------------------------------------------
%%%%%%%%%%%%%%%%%%%%%%%%%%%%%%%%%%%%%%%%%%%%%%%%%%%%%%%%%%%%%
\acknowledgments     %>>>> equivalent to \section*{ACKNOWLEDGMENTS}       

The authors wish to thank Dick Joyce at NOAO for pointers about the near-IR sky, Dr. James Graham of U.C. Berkeley for information about IRCAL.

\ni ShaneAO is made possible through an National Science Foundation Major Research Instrumentation grant, \#0923585.

%%%%%%%%%%%%%%%%%%%%%%%%%%%%%%%%%%%%%%%%%%%%%%%%%%%%%%%%%%%%%
%%%%% References %%%%%

%\bibliography{SPIEpaperSwimShARCS}   %>>>> bibliography data in refs.bib
%\bibliographystyle{spiebib}   %>>>> makes bibtex use spiebib.bst

\section{Appendices}
\subsection{Surface lists}
\label{ssec-surfs}

\subsubsection{Keck surface list}
Surface list for the Keck AO system:
\begin{verbatim}
Surface desc.           Material        Temp (C)
-------------           --------        --------
Primary                 Aluminum        2.6     
Secondary               Aluminum        2.6     
tertiary                Aluminum        2.6     

rotator M1              Aluminum        5       
rotator M2              Silver          5       
rotator M3              Aluminum        5       
Tip Tilt M              Silver          5       
OAP1                    Silver          5       
DM                      Silver          5       
OAP2                    Silver          5       
IR Dich 1 S1            tt_dichroic     5       
IR Dich 1 bulk          CAF2            5       
IR Dich 1 S2            NIR_AR          5       
Instr Window S1         NIR_AR          5       
Instr Window Bulk       CAF2            5       
\end{verbatim}

\subsubsection{Earlier Shane/IRCAL surface list}
\label{ssec-ircal-surfs}
Surface list for the existing Shane AO system with the IRCAL camera. Surfaces after the dashed line are within the dewar that contains the IRCAL camera, filter wheels etc.:
\begin{verbatim}
Surface desc.           Material        Temp (C)
-------------           --------        --------
Primary                 Aluminum        10      
Secondary               Aluminum        10      

1st Turning Mirror      Measured        10      
Tip/Tilt Mirror         Measured        10      
OAP 1                   Measured        10      
DM                      Measured        10      
OAP 2                   Measured        10      
NaDichroicSciencePath   NaR_splitter    10      
IR Mirror 1             Silver          10      
IR Mirror 2             Silver          10      
IRCAL window            Silver          10      
--------------------------------------------
Window                  Silver          -196    
Turn Mirror             Silver          -196    
OAP1                    Silver          -196    
Filter                  Silver          -196    
Cold Stop               Cold Stop       -196    
OAP2                    Silver          -196    
Turn Mirror             Silver          -196    
PICNIC                  PICNIC_QE       -196    
\end{verbatim}

\subsubsection{ShaneAO surface list}
\label{ssec-shaneao-surfs}
Surface list for the new ShaneAO system. Surfaces after the dashed line are within the dewar that contains the ShARCS camera, filter wheels etc.:
\begin{verbatim}
Surface desc.           Material        Temp (C)
-------------           --------        --------
Primary                 Aluminum        10      
Secondary               Aluminum        10      

Fold Mirror 1           Lick coating    10      
OAP 1                   Lick coating    10      
DM1                     Prot. Silver    10      
OAP 2                   Lick coating    10      
[TT dichroic            TT Dichroic     10  or  (LGS)
 Fold Mirror 2          Lick coating    10]     (NGS
OAP 3                   Lick coating    10      
MEMs window             MEMs_Window     10      
MEMs window             MEMs_Window     10      
MEMs DM                 Bare Gold       10      
MEMs window             MEMs_Window     10      
MEMs window             MEMs_Window     10      
NaDichroicSciencePath   WFS_Dichroic    10      
OAP 4                   Lick coating    10      
Fold mirror 3           Lick coating    10      
Dewar window            IR anti-refl    10      
-------------------------------------------
Dewar window            IR anti-refl    -196    
IR fold mirror          Ni+Gold         -196    
OAP_ircal_1             Ni+Gold         -196    
Filter                  J,H,K           -196    
[Grism                                  -196] - Spectroscopy
Cold stop               Cold stop       -196    
OAP_ircal_2             Ni+Gold         -196    
IR fold mirror          Ni+Gold         -196    
H2RG                    H2RG_QE         -196    
--------------------------------------------
\end{verbatim}

%\newpage
\subsection{ShARCS Detector Specifications}
\label{ss-sharcspec}

\begin{table}[h]
	\begin{center}	
		\begin{tabular}{|l|r|l|}
		\hline
		 Parameter & Value  & Units  \\
		\hline
		\hline
		 Pixel scale & 0.035 & arcsecs/pixel \\
		\hline
		 Pixel pitch  &   18 & microns \\
		\hline
		 Number pixels & 1024 & \\
		\hline
		 FOV          &   20 & arcsec square \\
		\hline
		 FOV diameter \footnotemark[1] 
		              &   28 & arcsecs \\
		\hline
	     Read noise   & 14.6 & e$^-$/CDS read \\
	     \hline
	     CDS Exp time & 10.6 & seconds \\
	     \hline
	     Fowler read noise \footnotemark[2]
	                  & 5.25 & e$^-$ \\
	     \hline
	     Full well    & 1.0\e{5} & e$^-$ \\
		\hline
		\end{tabular}
	\caption{ShARCS H2RG detector parameters}
	\label{tab-sharcspec}
	\end{center}
	\footnotemark[1]{circular aperture equivalent} \\
	\footnotemark[2]{32 samples} \\
\end{table}

\end{document}